\documentclass[aps,twocolumn,tightenlines]{revtex4}
\usepackage[dvipdf]{graphicx}
\begin{document}
\title{Index of refraction of gases for matter waves: \\
effect of the motion of the gas particles on the calculation of
the index.}

\author{ C. Champenois}
 \affiliation{Physique
des Interactions Ioniques et Mol\'eculaires (CNRS UMR 6633),
Universit\'e de Provence, Centre de Saint J\'er\^ome, Case C21,
13397 Marseille Cedex 20, France}

\author{ M. Jacquey}
\author{ S. Lepoutre}
\author{ M. B\"uchner}
\author{ G. Tr\'enec}
\author{ J. Vigu\'e}
\email{jacques.vigue@irsamc.ups-tlse.fr}
 \affiliation{ Laboratoire Collisions Agr\'egats R\'eactivit\'e UMR 5589,
\\ CNRS - Universit\'e de Toulouse-UPS, IRSAMC, Toulouse, France }

\date{\today}

\begin{abstract}

Two different formulae relating the index of refraction $n$ of
gases for atom waves to the scattering amplitude have been
published. We show here that these two formulae are not consistent
with the definition of the total scattering cross-section while
the formula developed by one of us (C.C.) in her thesis is in
agreement with this standard knowledge. We discuss this result, in
particular in the neutron case for which such an index was first
introduced. We finally evaluate the index of refraction as a
function of well known quantities and we discuss the order of
magnitude of the ratio of $(n-1)/n_t$, where $n_t$ is the gas
density.

\end{abstract}
\maketitle

\section{Introduction }

To describe the propagation of matter waves through a medium, it
is possible to use an index of refraction, as commonly done for
light. This idea was first introduced around 1940 for the
propagation of neutrons through matter (see the review papers by
L. L. Foldy \cite{foldy45} and by M. Lax \cite{lax51}).

Atom interferometry has permitted to study the propagation of an
atom wave through a dilute gas and the first measurements of the
index of refraction of gases for sodium waves were made in 1993 at
MIT by J. Schmiedmayer {\it et al.}
\cite{schmiedmayer93,schmiedmayer95} who measured the attenuation
and the phase shift of the transmitted wave. Further works in the
same laboratory have led to the observation of glory oscillations
of the index as a function of the sodium atom velocity
\cite{schmiedmayer97,hammond97,roberts02}. More recently, our
group has measured the index of refraction of several gases for
lithium waves \cite{jacquey07}.

Whatever is the nature of the wave and of the medium, the index of
refraction describes the modification of the propagation of an
incident wave due to the waves scattered in the forward direction
by the particles of the medium: the scattered waves interfere with
the incident wave and modify its phase and its amplitude. The
modification of the phase induces a modification of the wave
velocity, described by the real part of the index of refraction,
while the modification of the amplitude is described by its
imaginary part.

In practice, the index of refraction is proportional to the
complex forward scattering amplitude \cite{landau77,joachain75}.
The imaginary part of this amplitude is related to the total cross
section, which is traditionally measured by beam scattering
experiments, whereas its real part can be measured only by atom
interferometry. This amplitude exhibits resonances, for a
collision energy comparable to the potential well depth, and glory
oscillations, for larger energies. Glory oscillations are the
quantum consequence of the existence of an undeflected classical
trajectory due to the compensation of attractive and repulsive
forces \cite{pauly79}.

For atom waves, the forward scattering amplitude can be calculated
if the interaction potential between a particle of the wave and a
particle of the target gas is known. Several papers
\cite{vigue95,audouard95,audouard97,forrey96,champenois97,champenois99,forrey99,karchenko01,forrey02}
have discussed the theory of the index of refraction for atom
waves and their equations have been used to compare calculated
values of the index of refraction
\cite{audouard95,audouard97,forrey96,forrey97,champenois97,champenois99,leo00,blanchard03}
with experimental results. A detailed comparison is possible only
if the thermal motion of the target gas is not neglected and we
show here that the formulae \cite{forrey96,champenois97}
introduced to take into account this motion disagree with
collision physics. We propose a formula developed by one of us (C.
C.) in her thesis \cite{champenois99}: this formula agrees with
standard results of collision physics as well as with a recent
calculation based on quantum Boltzmann equation due to K.
Hornberger \cite{hornberger07}.

In the present paper, we first recall the previous results
concerning the index of refraction. Then, we explain why these
formulae are not in agreement with the Beer-Lambert law and we
extract from this discussion a new formula giving the imaginary
part of the index of refraction, which is generalized to the real
part of the index of refraction. We then discuss the differences
between the old and new formulae. We verify that our formula is in
agreement with the results concerning the index of refraction for
slow neutrons. Finally, we compare the order of magnitude of the
index of refraction of gases for light and atomic waves. In the
cases where the index of refraction for atom waves has been
measured, the values of the ratio $(n-1)/n_t$, $n_t$ being the
target gas density, are very close to the value of this ratio for
the index of refraction for ordinary gases for visible light. This
similarity is in fact a coincidence, without any physical meaning.

\section{Origin of the problem}

When one compares light waves and atom waves interacting with an
atomic (or molecular) target gas, there is a large difference,
which is precisely related to the motion of the target particles.

In the case of light, the photon velocity, almost equal to $c$ in
dilute matter, is considerably larger than the target particle
velocity. At the same time, the photon momentum is usually
considerably smaller than the target particle momentum. The atom
motion induces Doppler effect and the photon momentum induces atom
recoil. Because the velocity of light is usually so much larger
than the one of the target particles, these two effects have small
consequences on the index of refraction, if we except the
frequency range close to a sharp resonance line. This simple idea
remains true, even if the target particles move rapidly in the
laboratory, because one can always study the problem in their rest
frame.

In the case of atomic waves, the velocity of an atom of the wave
is usually comparable to the velocity of a target particle and, at
the same time, the scattering properties depend rapidly on the
relative velocity. Moreover, the momentum of an atom of the wave
and the one of a target particle are usually of comparable
magnitudes. In practice, it is absolutely necessary to take into
account the target particle motion to make a realistic calculation
of the index of refraction.

\section{Notations}

In the laboratory frame, a plane wave of wavevector ${\mathbf
k}_p$ describes the propagation in vacuum of a projectile $p$ of
mass $m_p$ and velocity ${\mathbf v}_p$:

\begin{equation}
\label{a0} \hbar {\mathbf k}_p = m_p {\mathbf v}_p
\end{equation}

\noindent The projectile can be any massive particle described by
quantum mechanics, a neutron, an electron, an atom or a molecule.
This wave propagates through a gas of density $n_t$ made of target
particles $t$ of mass $m_t$ and velocity ${\mathbf v}_t$, the
target particles being also described by quantum mechanics. The
wavevector $k_r$ describes the relative motion of the projectile
$p$ in the center of mass frame for a collision with a target
particle $t$: $k_r = \mu v_r/\hbar$, where $v_r = |{\mathbf v}_p -
{\mathbf v}_t|$ is the relative velocity and $\mu= m_p
m_t/(m_p+m_t)$ is the reduced mass. The present notations will be
used to write all previous formulae in order to facilitate
comparison.

\section{Index for fixed scattering centers}

When a plane wave of wavevector ${\mathbf k}_p$ enters in a medium
made of target particles $t$, its propagation is modified, with a
different wavevector ${\mathbf k}_{p,m}$ given by:

\begin{equation}
\label{a1} {\mathbf k}_{p,m} = n {\mathbf k}_p
\end{equation}

\noindent where $n$ is the index of refraction. If the medium is
described by a random distribution of fixed scattering centers,
the index of refraction $n$ is given by:

\begin{equation}
\label{a2}
 n = 1 + 2 \pi n_t \frac{f(k_p)}{k_p^2}
\end{equation}

\noindent where $f(k_p)$ is the forward scattering amplitude of
the wave scattered by one scattering center. $f(k_p)$ has the
dimension of a length and, as usual, the index $n$ is a
dimensionless quantity.  A general derivation of this formula is
found in the paper of L. Foldy \cite{foldy45} (see also the review
of M. Lax \cite{lax51} and references therein). Equation
(\ref{a2}) is the first order term of an expansion in powers of
the gas density $n_t$ and it is a good approximation if the
following conditions are fulfilled:

\begin{itemize}

\item  the mean distance $n_t^{-1/3}$ between nearest scattering
centers must be considerably larger than the projectile de Broglie
wavelength $\lambda_p =2\pi/k_p$, {\it i.e.} $n_t^{-1/3} \lambda_p
\ll 1$.

\item the mean distance $n_t^{-1/3}$ between scattering centers
must be considerably larger than the range of the interaction
potential. Unfortunately, as realistic atom-atom interaction
potentials $V(r)$ vanish only for an infinite distance $r$, the
range has not a clear definition for such potentials.

\item the density $n_t$ of scattering centers is low enough so
that the mean field correction is negligible {\it i.e.} $(n-1)\ll
1$. This last condition, which involves a condition on $k_p$ and
$f(k_p)$, depends on the interaction potential and collision
energy.
\end{itemize}

Practically, for thermal atoms waves and a target gas near room
temperature, with relative velocities of the order of $10^3$ m/s
and target densities up to  $n_t \approx 10^{19}$~m$^{-3}$ used in
the experiments
\cite{schmiedmayer93,schmiedmayer95,schmiedmayer97,hammond97,roberts02,jacquey07},
the mean interatomic distance $n_t^{-1/3}$ is larger than
$5\times10^{-7}$ m, the  index of refraction is of the order of
$|n-1| \lesssim 10^{-10}$ and these  three conditions are well
fulfilled.

In the early days of atom interferometry, the target particles
were treated as fixed scattering centers and the index was given
by equation (\ref{a2}): this was the case of the first paper
\cite{schmiedmayer93} dealing with the index of refraction for
atom waves, written by J. Schmiedmayer {\it et al.} in 1993 and of
the paper written by one of us (J.V.) in 1994 \cite{vigue95}. As
soon as experimental values \cite{schmiedmayer95} of the index of
refraction of gases for atomic waves became available in 1995, it
appeared necessary to take into account the target gas thermal
motion. The next section presents the equations used by different
groups.

\section{Formulae giving the index of a gas at thermal
equilibrium}

Three research groups have been involved in the calculation of the
index of refraction of gas for atomic waves and have worked on the
proper way to take into account the motion of the targets. For
convenience and clarity, the formulae used by each group are
presented separately.

\subsection{Publications of D. Pritchard and co-workers}
After a first paper where targets were considered at rest
\cite{schmiedmayer93}, this research group described in 1995 a set
of measurements of the index of refraction of gases for sodium
waves  \cite{schmiedmayer95}. To interpret their experiment, these
authors use the following equation:

\begin{equation}
\label{a3}
 n = 1 + 2 \pi n_t \frac{f(k_r)}{k_p k_r}
\end{equation}
\noindent where the thermal average is not explicitly discussed.
The same equation is also used in the review paper
\cite{schmiedmayer97} written by this group in 1997. A thermal
average is taken into account in the paper by T. D. Hammond  {\it
et al.} \cite{hammond97} published in 1997, with an index of
refraction given by:

\begin{equation}
\label{a4}
 n = 1 + 2 \pi \frac{n_t}{k_p}  \left\langle \frac{f(k_r)}{k_r}
\right\rangle
\end{equation}

\noindent where the brackets $\left\langle \right\rangle$ mean the
average over the velocity distribution of the target particles.

\subsection{Publications of A. Dalgarno and co-workers}

Equation (\ref{a4}) appears for the first time in the paper of R.
C. Forrey {\it et al.} \cite{forrey96}, written in 1995 and
published in 1996. In their paper, R. C. Forrey {\it et al.}
\cite{forrey96} calculate the distribution of the relative
wavevector $k_r$ and, for sake of completeness, we reproduce this
calculation in Appendix A. The group of A. Dalgarno has published
a series of papers on the index of refraction for atom waves
\cite{forrey97,karchenko01} or for electron waves \cite{forrey99}
and this work was continued by R. C. Forrey and co-workers
\cite{blanchard03}.

In 2002, R. C. Forrey {\it et al.} published a paper
\cite{forrey02} entitled 'On the statistical averaging procedure
for the refractive index of matter waves'. In this paper, they do
not give a complete derivation of their formula but they state
that a key step is the Lorentz invariance of the ratio $f(k_r)/k_r
= f(k_p)/k_p$.

\subsection{Publications of our research group in Toulouse}

In 1995, E. Audouard {\it et al.} published a calculation of the
index of refraction of argon gas for sodium waves
\cite{audouard95}. This work was the first one by our group taking
into account the effect of thermal averaging. We have made an
error in the algebra which was corrected in \cite{audouard97}. The
derivation of the thermal average formula was given in a following
paper by C. Champenois {\it et al.} \cite{champenois97}. Our
calculation was based on the Fizeau effect: this effect is a frame
dragging effect well known for light, which has also been studied
in the case of neutron matter waves
\cite{klein81,horne83,bonse86}. We had written:

\begin{equation}
\label{a5} {\mathbf k}_{p,m} = {\mathbf k}_{p} + \left\langle
(n_{CM}-1) {\mathbf k}_r \right\rangle
\end{equation}

\noindent with the center of mass index of refraction $n_{CM}$
given by:

\begin{equation}
\label{a6}  n_{CM}=1 + 2 \pi n_t  \frac{f(k_r)}{k_r^2}
\end{equation}

\noindent As above, the brackets $\left\langle \right\rangle$
stand for the average over the velocity distribution of the target
particles. We will not recall the final result \cite{champenois97}
corresponding to a Boltzmann distribution of the target gas,
because this result is not correct. After discussion with the
group of A. Dalgarno in 1998, whose results are presented above,
we were convinced that our formula was wrong, in particular
because our description of the Fizeau effect is not correct. In
her thesis \cite{champenois99}, Caroline Champenois derived a new
formula (see equations V.78 and V.81):

\begin{equation}
\label{a7}  n= 1 + 2 \pi n_t  \frac{m_p+ m_t}{m_t} \left\langle \
\frac{f(k_r)}{k_p^2} \right\rangle
\end{equation}

\noindent As the derivation of this formula was very involved, we
will not reproduce it here but, in the next section, we use a
simpler argument to convince the reader that this formula is the
right one.

\section{Disagreement with classic results and a new formula}

In this part, we show that neither equation (\ref{a4}) proposed by
R. C. Forrey {\it et al.} nor our equations (\ref{a5}) and
(\ref{a6}) are in agreement with well accepted atomic collision
results. We first introduce the total scattering cross section and
relate it to the imaginary part of the index of refraction.

\subsection{The Beer-Lambert law}

In nonrelativistic mechanics, the total scattering cross section
$\sigma(v_r)$ is related to the number of collisions $dN_{coll}$
occurring during a time $d\tau$ in a volume $dV$ between a
projectile $p$ and a target $t$:

\begin{equation} \label{f1}
 \frac{dN_{coll}}{d\tau dV} = n_p n_t \sigma (v_r) v_r
 \end{equation}

\noindent From this equation, we deduce the mean number of
collisions  $dN_{coll, p}/d\tau$ encountered by a projectile $p$
per unit time:

\begin{equation} \label{f2}
 \frac{dN_{coll, p}}{d\tau} = n_t \sigma (v_r) v_r
\end{equation}

\noindent We now consider a beam of projectiles $p$ crossing a
slab of target gas, with the velocity ${\mathbf v}_p$
perpendicular to the slab. A slab of thickness $dL$ is crossed by
a projectile $p$ in a time $d\tau= dL/v_p$ and the mean number of
collisions for a projectile is given by:

\begin{equation} \label{f3} dN_{coll, p} = n_t \sigma (v_r)
\frac{v_r}{v_p} dL \end{equation}

\noindent From this equation, one can deduce the transmission $T$
of the slab {\it i.e.} the fraction of the incoming flux which has
crossed the slab without any collision. For a finite thickness
$L$, the transmission $T$ is obtained by a straightforward
integration:

\begin{equation} \label{f4} T = \exp \left[- n_t \sigma (v_r)
\frac{v_r}{v_p} L \right] \end{equation}

\noindent This equation is the Beer-Lambert law, usually written
with an effective cross section $\sigma_{eff}(v_p) = \sigma (v_r)
v_r/v_p $:

\begin{equation}
\label{f41} T = \exp \left[- n_t \sigma_{eff}(v_p) L  \right]
\end{equation}

\noindent If the target velocity is spread, with a normalized
distribution $P({\mathbf v}_t)$ ({\it i.e.} verifying $\int
P({\mathbf v}_t) d^3{\mathbf v}_t=1$), we must replace the
effective cross section $\sigma_{eff}(v_p)$ in equation
(\ref{f41}) by its average $\left\langle
\sigma_{eff}(v_p)\right\rangle$ given by:

\begin{equation} \label{f5}
\left\langle \sigma_{eff}(v_p)\right\rangle =\int P({\mathbf v}_t)
\sigma (v_r) \frac{v_r}{v_p} d^3{\mathbf v}_t. \end{equation}

\subsection{Wave-like description of the attenuation of a beam by
a slab}

We are going to calculate the transmission $T$ of the beam through
the same slab of length $L$, using the wave point of view. Let
$\psi_{inc}$ be the incident wave and $\psi_{trans}$ the
transmitted wave given by:

\begin{equation}
\label{f8} \psi_{trans} =   e^{i \left(n-1\right)k_p L} \psi_{inc}
= te^{i \varphi}\psi_{inc}
\end{equation}

\noindent with $t = \exp\left[-{\mathcal{I}}m(n-1) k L\right]$ and
$\varphi = {\mathcal{R}}e(n-1)k L$. The transmission in intensity
is $T = t^2$, which depends solely on the imaginary part of
$(n-1)$:

\begin{equation} \label{f9} T = \exp \left[- 2 {\mathcal{I}}m(n-1) k_p L
\right] \end{equation}

\subsection{Consequences for the index of refraction}

ASs the attenuation of the beam calculated in the two formalisms
must be the same, the imaginary part of the index of refraction is
related to the effective cross section by the following equation:

\begin{equation}
\label{f10}    {\mathcal{I}}m(n-1) = \frac{n_t \langle
\sigma_{eff}(v_p) \rangle}{2k_p}
\end{equation}

\noindent The total cross section is related by the optical
theorem to forward scattering amplitude \cite{joachain75}:

\begin{equation}  \label{f11} \sigma(v_r) =4 \pi \frac{\mathcal{I}m(f(k_r))}{k_r}
\end{equation}

\noindent Using this relation and equations (\ref{f5}) and
(\ref{f10}), we obtain a formula giving the imaginary part of the
index of refraction in agrement with Beer-Lambert law:

\begin{equation}
\label{f12}   {\mathcal{I}}m(n-1) = 2 \pi n_t \frac{m_p+ m_t}{m_t}
\left\langle \frac{{\mathcal{I}}m(f(k_r))}{k_p^2} \right\rangle
\end{equation}

\noindent where the brackets $\left\langle \right\rangle$ mean the
average over the target velocity distribution $P({\mathbf v}_t) $.
Once we have an expression for the imaginary part of $(n-1)$, we
get the real part by a simple generalization:

\begin{equation}
\label{f13}   n= 1 + 2 \pi n_t \frac{m_p+ m_t}{m_t} \left\langle
\frac{f(k_r)}{k_p^2} \right\rangle
\end{equation}

\noindent This equation agrees with the result derived by C.
Champenois in her thesis \cite{champenois99}. It agrees also with
a result recently obtained by K. Hornberger using the formalism of
the quantum Boltzmann equation \cite{hornberger07}. We want to
point out that equation (\ref{f13}) can be applied with any type
of velocity distributions, as could be produced by a gas flow or a
molecular beam, and not only with a Maxwell-Boltzmann distribution
corresponding to thermal equilibrium.

\subsection{Comparison of the new and previous formulae}

The new formula, equation (\ref{f13}), differs from equation
(\ref{a4}) established by R. C. Forrey {\it et al.}
\cite{forrey96,forrey02}:

- the denominator is $k_rk_p$ in equation (\ref{a4}) and $k_p^2$
in equation (\ref{f13});

- the mass ratio $(m_p+ m_t)/m_t$ present in equation (\ref{f13})
is absent from equation (\ref{a4}).

- from equation (\ref{a4}), we deduce an effective cross section
given by:

\begin{equation} \label{b1} \langle\sigma_{eff}(v_p) \rangle
= \int P({\mathbf v}_t)\sigma(v_r) d{\mathbf v}_t \end{equation}

\noindent This expression of $\langle\sigma_{eff}(v_p) \rangle$
differs from equation (\ref{f5}) and in the $v_p\ll \alpha $
limit, where $\alpha$ is the thermal velocity defined in Appendix
A, the effective cross section given by equation (\ref{b1}) is
independent of the projectile velocity $v_p$, while the correct
behavior given by equation (\ref{f5}) is $\langle\sigma_{eff}(v_p)
\rangle \propto 1/v_p$. This is a well known result, recognized in
everyday's life: we run under the rain to get less wet!

However, in the opposite limit when the target gas temperature
vanishes, ${\mathbf v}_t = {\mathbf 0}$ so that $v_r= v_p$ and
$k_r= \mu v_p$, it is easy to verify that equation (\ref{f13}) and
equation (\ref{a4}) are then equivalent. This equivalence suggests
that, even when the temperature does not vanish, the index
calculated by these two formulae will not differ strongly as long
as $\alpha \ll v_p$.

\section{Discussion of the neutron case}

It is also interesting to apply equation (\ref{f13}) to the well
known case of neutron waves. Neutrons are scattered only by
nuclei, if we except the case of ferromagnetic materials in which
the magnetic interactions of the neutron spin cannot be neglected.
In the low energy domain where the formalism of the index of
refraction is useful, the neutron-nucleus scattering process is
almost always dominated by s-wave scattering
\cite{sears78,rauch00}.

The index of refraction is frequently calculated as the
consequence of an effective potential which is related to the
scattering length of the neutron-nucleus interaction potential.
However, there are few papers devoted to the theoretical relation
between scattering theory and the index of refraction for
neutrons, in which the motion of the nuclei is taken into account.
This is the case of the papers by B. A. Lippmann and J. Schwinger
\cite{lippmann50} and also of a brief note by D. Kleinman and G.
Snow \cite{kleinman51}, who state: ``This derivation of the index
clearly shows that there is no Doppler effect due to the motion of
the nuclei, because the $\lambda$ in the formula is the neutron
wavelength relative to the boundary of the slab''.

Obviously, the neutron case differs from the case of atom waves
only by the fact that s-wave ($l=0$) scattering dominates the
forward scattering amplitude, which, in this case, is given by:

\begin{eqnarray}
\label{n1}   f(k_r) &=& \frac{\exp\left(i\delta_0\right)
\sin\delta_0}{k_r} \nonumber \\ &\approx& -a (1-ik_r a)
\end{eqnarray}

\noindent where $\delta_0$ is the s-wave phase shift, $a$ is the
scattering length defined by $a= -\mbox{lim } (\tan \delta_0/k_r)$
when $k_r \rightarrow 0$. If we keep only the leading term in
equation (\ref{n1}), $f(k_r) \approx -a$, our formula (\ref{f13}),
is equivalent to the result of D. Kleinman and G. Snow
\cite{kleinman51}:
\begin{equation}
\label{neutrons}   n= 1 -2 \pi n_t \frac{m_p+m_t}{m_t}
\frac{a}{k_p^2}
\end{equation}

\noindent As $k_r$ is absent from the result, there is no Doppler
effect on the index of refraction. However, if we take into
account the first order term in $k_r$ in equation (\ref{n1}), the
forward scattering amplitude has a non-vanishing imaginary part,
$\mathcal{I}m \left(f(k_r)\right) \approx k_r a^2$. As this
imaginary part is linear in $k_r$, the imaginary part of the index
of refraction is sensitive to the motion of the target particles
but this imaginary part, which is very small, is usually ignored.
The $k_r$-dependence of the imaginary part of the index of
refraction has no practical consequences but this remark proves
that the absence of Doppler effect on the index of refraction of
matter for neutrons is a very special property valid only for the
real part of the index of refraction in the s-wave limit.

\section{Comparison of the order of magnitude of the index of refraction of gases
for light and matter waves}

Up to now, we have not discussed the numerical value of the index
of refraction of gases for matter waves. A somewhat surprising
feature is that the index of refraction of gases for matter waves
\cite{schmiedmayer95,jacquey07} and the index of refraction of
transparent gases for light has similar values, when the gas
density $n_t$ is the same. In this section, we calculate the value
of the $(n-1)/n_t$ ratio from first principles in both cases and
we compare the results, in order to understand the origin of this
similarity.

J. Schmiedmayer {\it et al.} \cite{schmiedmayer95} give the values
of the real and imaginary parts of $(n-1)$ of several gases at
$T=300$ K for sodium matter waves with a velocity $v_p= 1000$ m/s.
All the values of $\mathcal{R}e(n-1)$ or of $\mathcal{I}m(n-1)$
are in the range $(0.14-2.49) \times 10^{-10}$ for a gas pressure
equal to $1$ mTorr at $T=300$ K. This pressure corresponds to a
gas density $n_t \approx 3.2 \times 10^{19}$ m$^{-3}$ from which
we find that the ratio $\mathcal{R}e(n-1)/n_t$ of
$\mathcal{I}m(n-1)/n_t$ is in the range $(0.4 - 8)\times 10^{-30}$
m$^3$. Our results for lithium waves \cite{jacquey07} are somewhat
larger, in the range $1-2\times 10^{-29}$ m$^3$.

It is well known that the index of refraction of air for visible
light is almost purely real with its value given by $(n-1) \approx
2.8 \times 10^{-4}$ at ordinary pressure and a temperature of
$288$ K, with a density $n_t \approx 2.55 \times 10^{25}$
m$^{-3}$. From these values, we calculate a ratio $(n-1)/n_t
\approx 1.1 \times 10^{-29}$ m$^3$ for air. Similar values will be
obtained for other transparent gases.

The ratio $(n-1)/n_t$ has comparable values for light or matter
waves, although the only common feature is that they involve an
index of refraction. We now evaluate these two ratio from first
principles.

In optics, the index of refraction of an atomic or molecular gas
is dominated by the electric dipole transitions, with its value
given by:
\begin{equation}
\label{m1} n(\omega) =\left[1+4 \pi n_t \alpha(\omega)
\right]^{1/2} \approx 1 + 2 \pi n_t \alpha(\omega)
\end{equation}
\noindent where $n_t$ is the gas density, $\alpha(\omega)$ the
atomic or molecular electric polarizability for an angular
frequency $\omega$. In the visible region of the spectrum, where
$\omega$ is usually smaller than the angular frequency of the main
resonance transitions, $\alpha(\omega)$ is close to its static
value $\alpha(0)$. We now use atomic units, with $\alpha(0) =
a_0^3 \alpha_{au}(0)$, where $a_0$ is the Bohr radius ($a_0
\approx 0.529 \times 10^{-10}$ m) and we get:

\begin{equation}
\label{m2} \frac{n(\omega)-1}{n_t} \approx 2 \pi a_0^3
\alpha_{ua}(0)
\end{equation}
\noindent Tabulated values of $\alpha_{au}$ for atoms
\cite{miller77} vary from $1.4$ a.u. for Helium up to $450$ a.u.
for Francium and the polarizability of small molecules has similar
values (for instance, nitrogen dimer N$_2$ has a polarizability
$\alpha_{ua}(0)\approx 11.9 $ \cite{kramer70}).

For matter waves, the index of refraction is given by equation
(\ref{f13}). The imaginary part $\mathcal{I}m(n-1)$ is related to
the total cross-section $\sigma(v_r)$ and, for a purely attractive
$-C_6/r^{-6}$ potential, there is a closed form expression of
$\sigma(v_r)$ \cite{landau77}:

\begin{equation}
\label{m3} \sigma(v_r) = 8.08 \left[\frac{C_6}{\hbar
v_r}\right]^{2/5}
\end{equation}

\noindent The $8.08$ factor is the numerical value of a
complicated expression involving gamma function. This result is
valid in an intermediate range of energy, with many partial waves
contributing to the scattering amplitude \cite{pauly79}. This
result does not explain the glory oscillations, which exists only
when the potential is attractive at long range and repulsive at
short range, but this result gives the correct value of the
cross-section averaged over the glory oscillations. As we want
only an order of magnitude, we will make some simplifications, by
neglecting the thermal motion of the target particle and by
assuming that $m_p \ll m_t$, so that we can replace $v_r$ by
$v_p$. We thus get an expression of the imaginary part of
$(n-1)/n_t$:

\begin{equation}
\label{m4} \frac{\mathcal{I}m(n-1)}{n_t}  = \frac{4.04}{k_p}
\left[\frac{C_6}{\hbar v_p}\right]^{2/5}
\end{equation}
\noindent For a purely attractive $-C_6/r^{-6}$ potential, the
real and imaginary parts have similar values: Schmiedmayer {\it
{\it et al.}} \cite{schmiedmayer95} calculated the ratio $\rho =
{\mathcal{R}}e(n-1) / {\mathcal{I}}m(n-1)$ and found that it is
constant and equal to $ \rho= 0.7265$ (see also
\cite{champenois97}). We can now express $\mathcal{I}m(n-1)/n_t$,
using atomic units for $C_6 = C_{6au} m_ec^2\alpha_{fs}^2a_0^6$
($m_e$ is the electron mass and $\alpha_{fs} \approx 1/137.037$ is
the fine-structure constant). We get:

\begin{equation}
\label{m5} \frac{\mathcal{I}m(n-1)}{n_t}  =  4.12\times10^{-3}
 a_0^3 \left(\frac{m_e}{m_p}\right)\left(\frac{c}{v_p}\right)^{7/5}\left(C_{6ua}\right)^{2/5}
\end{equation}

\noindent Tabulated values of the atom-atom $C_{6au}$ coefficient
span the range from $1.47$ for Helium up to ca. $7260$ a.u. for
Cesium \cite{kramer70,tang76} (we quote these references which
provide a large set of $C_6$ values). The comparison of equation
(\ref{m2}) and (\ref{m5}) proves that, the index of refraction of
gases for light and matter waves have very different expressions
and, if they have comparable values as in the cases discussed
above, this must be considered as a pure coincidence. The index
for matter waves has a rapid velocity dependence in $v_p^{-7/5}$
in a large velocity range and it would be considerably larger for
lower projectile velocities, provided that the target gas velocity
can still be neglected.

\section{Conclusion}

In this paper, we have shown that the equations proposed by R. C.
Forrey {\it et al.} \cite{forrey96,forrey02} and by our group
\cite{champenois97} to take into account the motion of the gas
particles in the calculation of the index of refraction for matter
waves are not consistent with the traditional definition of a
cross-section. Following the result obtained by C. Champenois in
her thesis \cite{champenois99}, we propose a new formula for the
index of refraction:

\begin{equation}
\label{c1}   n = 1 + 2 \pi n_t \frac{m_p+ m_t}{m_t} \left\langle
\frac{f(k_r)}{k_p^2} \right\rangle
\end{equation}

\noindent This formula is consistent with classic results of
collision theory and it is also in agreement with a result
recently obtained by K. Hornberger using the formalism of the
quantum Boltzmann equation \cite{hornberger07}. We have used this
formula when comparing our measurement of the index of refraction
of gases for lithium waves with theoretical values
\cite{jacquey07}. Equation (\ref{c1}) agrees also with the formula
giving the index of refraction of matter for neutrons.

Finally, in the limiting case where the thermal velocity $\alpha$
of the gas is considerably smaller than the projectile velocity
$v_p$, equation (\ref{c1}) and equation (\ref{a4}) proposed by R.
C. Forrey {\it et al.} \cite{forrey96,forrey02} are equivalent but
they differ considerably in the opposite limit $v_p \ll \alpha$.

\section{Acknowledgements}

We thank K. Hornberger for the communication of his recent results
prior to publication and for a very stimulating discussion. We
also thank A. Dalgarno for inviting C. Champenois for one month in
1998 and for the interaction between our research groups at that
time.

\section{Appendix A}

We calculate the distribution $P(v_r)$ of the relative velocity
$v_r$, starting from a normalized Maxwell-Boltzmann distribution
of the target velocity ${\mathbf v}_t$:

\begin{equation}
\label{maxwell} P_{MB}({\mathbf v}_t)d^3{\mathbf v}_t =
\frac{1}{\pi^{3/2}\alpha^3} \exp\left( -{\mathbf
v}_t^2/\alpha^2\right)d^3{\mathbf v}_t
\end{equation}

\noindent with $\alpha = \sqrt{2 k_BT/m_t}$, $k_B$ being the
Boltzmann constant and $T$ the temperature. Using ${\mathbf v}_t =
{\mathbf v}_p -{\mathbf v}_r$, we can write:

\begin{equation}
\label{app3} P({\mathbf v}_r) d^3{\mathbf v}_r =
\frac{1}{\pi^{3/2}\alpha^3} \exp\left[ -({\mathbf v}_p - {\mathbf
v}_r)^2/\alpha^2\right] d^3{\mathbf v}_r
\end{equation}

\noindent We take the $z$ axis along to the projectile velocity
${\mathbf v}_p$, we use spherical coordinates for ${\mathbf v}_r$,
$v_r$ being its modulus,  $\theta$ its angle with the $z$-axis
$\theta$ and and  $\varphi$ its azimuth. Integration over
$\varphi$ and $\theta$ is easy and we get:

\begin{equation}
\label{app4} P(v_r) d v_r=  \frac{2 v_r}{\pi^{1/2}\alpha v_p}
\exp\left[ -\frac{v_p^2 + v_r^2}{\alpha^2}\right]\sinh \left[
\frac{2 v_p v_r}{ \alpha^2} \right] d v_r
\end{equation}
\noindent which is normalized, $\int_0^{\infty} P(v_r) d v_r=1$.

\end{document}